\documentclass[runningheads]{cl2emult}

\newcommand{\be}{\begin{equation}}
\newcommand{\ee}{\end{equation}}
\newcommand{\bea}{\begin{eqnarray}}
\newcommand{\eea}{\end{eqnarray}}

\begin{document}
\title*{The Meaning of Decoherence\thanks{
To be published in the proceedings of the Bielefeld conference on
``Decoherence: Theoretical, Experimental, and Conceptual Problems",
edited by P. Blanchard, D. Giulini, E. Joos, C. Kiefer, and I.-O.
Stamatescu (Springer 1999).
This contribution is based on Sects.\ts 4.2 - 4.4 of  Zeh (1999).
}}
\toctitle{The Meaning of Decoherence}
\titlerunning{Meaning of Decoherence}
\author{H. D. Zeh}
\authorrunning{H. D. Zeh}
\institute{Institut f\"ur Theoretische Physik, Universit\"at Heidelberg,
D-69120 Heidelberg, Germany}

\maketitle              

\begin{abstract}
The conceptual and dynamical aspects of decoherence are analyzed, while
their consequences on several fundamental applications are discussed.
This mechanism, which is based on a universal Schr\"odinger
equation, is furthermore compared with the phenomenological description
of open systems by means of `quantum dynamical maps'.
\end{abstract}

\section{Ensembles, Entanglement, and Zwanzig Projections}
Decoherence is usually defined
as the practically irreversible disappearance of
certain nondiagonal elements from the density matrix of a
bounded but open system. It can be explained by the latter's
unavoidable interaction with its environment according to  the
Schr\"odinger equation for a global system under certain initial
conditions. This {\em delocalization of phase relations} seems to form
the most ubiquitous irreversible process in nature --- similar to (but
in general far more efficient than) the Boltzmann equation in classical
statistical physics. In order to understand its importance correctly,
one has to analyze carefully the meaning of the quantum
mechanical {\em density matrix}, since its usual introduction by means
of `quantization rules' would be insufficient for this purpose. As will
be explained below, there are indeed two quite {\em different}
interpretations of the density matrix. Their confusion would lead to a
common `naive' misinterpretation of the concept of decoherence as
describing a collapse of the wave function.

If the wave function (that is, the quantum mechanical {\em state}) of a
physical system is assumed to be physically defined, but not completely
known, one may often describe this incomplete knowledge by an {\em
ensemble} consisting of wave functions $\psi_\alpha$ with
probabilities
$p_\alpha$.  In this ensemble, the probabilities
$p_\alpha $ (rather than a density matrix $\rho (q,q^\prime )$  that
would arise from the formal quantization rules) form the analog of the
classical probability  distribution on phase space,
$\rho (p,q)$. The meaning of the density matrix can only be
appreciated when considering {\em ensemble expectation values} of
observables
$A$, that is, {\em mean values} of expectation values with respect to
the various states $\psi_\alpha $ which form the ensemble:
$$
\langle A \rangle := \sum_\alpha p_\alpha \langle  \psi_\alpha |A|
\psi_\alpha
\rangle
= {\rm Trace}\{A\rho \} = \sum_n a_n \langle \phi_n |\rho| \phi_n \rangle
\quad , \eqno(1a)
$$
with
$$
\rho := \sum_\alpha  |\psi_\alpha \rangle
p_\alpha \langle \psi_\alpha |
\quad {\rm and} \quad A := \sum_n |\phi_n \rangle a_n \langle \phi_n|
 \quad. \eqno(1b) \setcounter{equation}{1}
$$
The symbol $\langle A\rangle $ denotes here  a twofold mean:
with respect to the ensemble of quantum states $\psi_\alpha$ with their
probabilities
$p_\alpha$, {\em and} with respect to the quantum  mechanical
indeterminism of measurement results
$a_n$ with their probabilities $| \langle \phi_n |
\psi_\alpha \rangle |^2 $, valid for {\em given}  quantum states
$\psi_\alpha$. In this way, the concept of a  density matrix  depends on
the probability interpretation  of the wave
function ---  although {\em not} on any specific kinematical  concept
(such as hidden or classical variables) that would characterize the
objects to which these probabilities apply, or where and how
they might dynamically arise.

This ensemble interpretation of the density matrix
according to $\rho =
\break
\sum_\alpha  |\psi_\alpha \rangle  p_\alpha \langle \psi_\alpha |$ does
not require the members $\psi_\alpha$ of the ensemble of wave
functions to be mutually orthogonal. They may in general even
form an overcomplete set. Therefore,
the ensemble cannot be uniquely recovered from the density matrix. Von
Neumann's entropy,
\be
 S := -k {\rm Trace}\{\rho \ln \rho\} \quad ,
\ee
would represent an ensemble
entropy in the form
$ -k\sum p_\alpha \ln p_\alpha$ only for the special
ensemble consisting of the orthonormal
eigenstates of
$\rho$. Its conservation
describes dynamical determinism, provided the inner products  between the
states
$\psi_\alpha$ are also
conserved. This  requires the
{\em unitarity} of the Schr\"odinger equation, since the formal density
matrix does not distinguish between norm and probability of a wave
function.

The mapping of ensembles of wave functions onto those which diagonalize
their density matrix is an idempotent, information-reducing operation.
Nonetheless, one may derive the von Neumann equation,
\be
i {\partial \rho \over \partial t} = [H,\rho ] =: \hat L\rho \quad ,
\ee
 from the ensemble interpretation and the
further assumption that all wave functions defining the ensemble satisfy
a Schr\"odinger equation $i\partial
\psi_\alpha/\partial t = H\psi_\alpha$ with one and the {\em same
Hamiltonian}
$H$. Although similar assumptions are used in classical  statistical
mechanics, presuming the Hamiltonian to be `given' does not appear
as an entirely consistent procedure while regarding {\em states} as
incompletely known: the exact Hamiltonian of a bounded classical system
would in general essentially depend on the uncontrollable state of its
environment.

Rather than describing an ensemble of wave functions, a density matrix  may
also represent the local or `reduced' perspective of {\em entangled}
quantum systems. The generic quantum state of any two combined systems
(with variables
$x$ and $y$, say) may be written as
\be
\psi(x,y) = \sum_{m,n}c_{mn}\phi_m(x) \Phi_n(y) \quad .
\ee
For spatially distinct subsystems, this entanglement represents the
fundamental (kinematical) {\em quantum nonlocality}.  All {\em partial}
measurements at one of the subsystems ($\phi(x)$, say) can then be
characterized by expectation values of  observables $A_\phi$,
\be
\langle A_{\phi} \rangle := {\rm Trace}\{A_{\phi}\rho_{{\rm total}} \} =
{\rm Trace}_\phi \{ A_{\phi}\rho_{\phi} \} \quad .
\ee
Here, $\rho_\phi$ is defined as a partial trace, $\rho_\phi :=
{\rm Trace}_\Phi\{ \rho_{{\rm total}}\} $, while
$\rho_{{\rm total}}$ may still represent a wave function (or pure state),
$\rho_{{\rm total}} :=    |\psi \rangle  \langle \psi|$. This {\em new}
density matrix
$\rho_\phi$ is explicitly defined by the expansion
coefficients $c_{mn}$ of the total state (4),
\be
(\rho_\phi)_{mm^\prime} := \langle \phi_m|\rho_\phi|\phi_{m^\prime}
\rangle =
\sum_n c_{mn} c_{m^\prime n}^\ast
\quad ,
\ee
rather than by a statistical distribution $p_\alpha$ according to
$\sum_{\alpha} c_{\alpha m} p_\alpha c^\ast_{\alpha m'}.$ Although  it
cannot be distinguished by local {\em operations} from the density
matrix describing an ensemble of quantum states, it does here
evidently {\em not} represent such an ensemble. Mere incomplete
information would mean that {\it one} definite member of the ensemble
described reality. Therefore, this `apparent ensemble' or  `improper
mixture' (d'Espa\-gnat 1966) must not be used to {\em explain} the
probability interpretation on that it has been based in (5). The
fundamental difference between proper and improper mixtures cannot be
overcome (though possibly obscured) by applying the formal limit of an
infinity of degrees of freedom (cf. Hepp 1972). Statistical operators
for all subsystems (as widely used in the phenomenological quantum
theory of open systems) are therefore insufficient, as they neglect
nonlocal quantum correlations. This formalism remains blind to the
measurement problem (see below).

Any density matrix, such as $\rho_\phi$ or
$\rho_\Phi := {\rm Trace}_\phi \{\rho_{{\rm total}}\}$,  is hermitian, and
can therefore be diagonalized in the form
$\rho_\phi =
\sum_n |\tilde
\phi_n
\rangle p_n \langle \tilde \phi_n |$ that defines its eigenbasis
$\{ \tilde \phi_n \}$. This form represents an (in general apparent)
ensemble of {\em orthogonal} states. By using this diagonal form and
(6) one observes that all eigenvalues $p_n$ are non-negative.
Phenomenological {\em dynamical maps} (Sect.\ts 3)
must  therefore be chosen `completely positive' by hand, that is, they
have to conserve this property for {\em all} density matrices (including
those of their subsystems).

For an entangled state such as (4), the eigenbasis of
both subsystem density matrices defines the {\em Schmidt  canonical form},
\be
\psi(x,y) = \sum_k \sqrt {p_k} \tilde \phi_k(x) \tilde \Phi_k(y) \quad ,
\ee
which, in contrast to  (4), is a
single sum (Schmidt 1907, Schr\"odinger 1935). For given subsystems,
this representation (and therefore also its time dependence --- see
K\"ubler and Zeh 1973) is completely defined by the state
$\psi$ of the total system (except for degenerate probabilities).
All phase factors which could multiply the roots of the
formal probabilities
$\sqrt{p_k}$ in (7) have here been absorbed into the orthonormal
states
$\tilde
\phi_k$ or
$\tilde
\Phi_k$.
 Since indistinguishable particles cannot be used  to define
subsystems, entanglement is not understood to include
the formal correlations which describe symmetrization or
antisymmetrization of the wave function. These pseudo-correlations are
merely an artifact from using classical particle concepts.

Complete {\em neglect} of all correlations between two
subsystems can be formally described by a {\em Zwanzig projection} (an
idempotent mapping of density matrices),
\be
 \hat P_{\rm sep} \rho := \rho_{\phi}
\otimes
\rho_{\Phi}
\quad .
\ee
Such operators $\hat P$ on density matrices, with $\hat P^2 = \hat P$,
form a convenient tool in deriving master equations for the `relevant'
part
$\rho_{{\rm rel}} := \hat P
\rho$ of
$\rho$, such as
\be
\left\{ \partial \rho_{\rm rel}  \over \partial t \right\}_{\rm master}
:= {\hat P {\rm e}^{-i \hat L \Delta t} \rho_{\rm rel} - \rho_{\rm rel}
\over
\Delta t}
\ee
They {\em reduce} information contained in the density
matrix (and thus raise the entropy) if they are genuine projections
(linear and hermitian --- see Sect.\ts3). The stronger
Zwanzig projection of locality,
$\hat P_{\rm local}\rho = \prod _k
\rho_{\Delta V_k}$, leading to a density matrix that  factorizes with
respect to small volume elements $\Delta V_k$ in space, would be
required in order to arrive at the approximate concept of an {\em entropy
density}
$s({\vec r})$. In quantum mechanics, a Zwanzig projection that neglects
certain interference terms,
\be
\hat P_{\rm semidiag} \rho := \sum_n
P_n
\rho P_n
\quad ,
\ee
is often useful. Here, $\{P_n\}$ is a complete set of projectors on
mutually orthogonal Hilbert subspaces. A master equation
requires that the irrelevant part,
$\rho_{\rm irrel} := (1-\hat P) \rho$, is {\em dynamically} irrelevant
in the future (similar to the particle correlations neglected in
Boltzmann's {\em Sto\ss zahl\-ansatz}).

A reduction of information less than by  $\hat
P_{\rm sep}$ is obtained by the
relevance concept of {\em
classical  correlations only},
\be
\hat P_{\rm classical} ( |\psi
\rangle
\langle
\psi |) := \sum_k p_k |\tilde \phi_k\rangle \langle \tilde \phi_k| \otimes
|\tilde \Phi_k\rangle \langle \tilde \Phi_k| \quad ,
\ee
here again written in the Schmidt-canonical basis. The Zwanzig projection
$\hat P_{\rm classical} $
 retains all `statistical'  correlations (based on
incomplete information) while dropping all  quantum
correlations
(entanglement). In the Schmidt
representation the latter would require a twofold sum over $k$ and
$k^\prime$ in the density matrix (in an arbitrary representation  a
sum over {\em two pairs} of indices). The rhs of (11) can
always be interpreted as representing lacking
information about one `real' product state. The
presence of genuine quantum entanglement can be
recognized in {\em any} representation of the density matrix (Werner
1989, Peres 1996).

A reduced density matrix does in general {\em not} obey a von Neumann
equation any more if the total wave function $\psi$ evolves according to a
Schr\"odinger equation. Its dynamics cannot be autonomous. Although it can
be explicitly formulated (K\"ubler and Zeh 1973, Pearle 1979), its solution
would in general require solving the Schr\"odinger equation for the total
system. Indeed,
$\rho_\phi$ multiplied by the unit operator in
$\Phi$-subspace represents another Zwanzig
projection,
\be
\hat P_{\rm sub} \rho_{{\rm total} }
:=
\rho_\phi\otimes 1_\Phi
\quad .
\ee
Phenomenological master equations for
$\rho_\phi$  are referred to as `open systems quantum dynamics'
(see Sect.\ts3). They are often derived by assuming a heat bath as an
(uncorrelated) environment (Favre and Martin 1968, Davies 1976). However,
from a fundamental point of view, master equations should {\em explain}
the presence of heat baths (that is, canonical ensembles described by a
temperature parameter)  rather than presuming
their existence.

The expectation values (1a) and (5), which led to the concept of a
density matrix, refer to  probabilities for the outcomes of quantum
measurements. Von Neumann (1932) introduced his dynamical concept of {\em
ideal measurements} (or {\em measurements
of the first kind}) as unitary interactions   between microscopic systems
and measurement devices. They represent {\em
forks of causality} (spreading information
--- see Zeh 1999), defined by the transition
$\phi_n\Phi_0
\mathop{\rightarrow}\limits_t
\phi_n\Phi_n$, where $\phi_n$ is an eigenstate
of an observable $A = \sum_n|\phi_n \rangle a_n \langle \phi_n
|$.
$\Phi_0$ is the  initial state of the apparatus, and $\Phi_n$
its `pointer position' corresponding to the result $n$. The observable
$A$ is {\em defined by this interaction between wave functions} up to the
scale factor
$a_n$ if the states
$\phi_n$ are orthogonal. If the microscopic system is
initially in one of the eigenstates, it does not change during an ideal
measurement, while the apparatus evolves into the corresponding pointer
state
$\Phi_n$.  (Non-ideal measurements could be similarly described by replacing
$\phi_n$ with a different final state
$\phi_n^\prime $.)

However, for a general initial state, $\sum_n
c_n\phi_n$, one obtains for the same interaction and for  the same
initial state of the apparatus
\be
 \left( \sum_n c_n \phi_n \right) \Phi_0 \quad
\mathop{\rightarrow}\limits_t \quad \sum_n c_n \phi_n \Phi_n =: \psi
\quad .
\ee
The rhs is an {\em entangled} state, while an {\em ensemble} of different
measurement results (that is, of states $\phi_n\Phi_n$ with probabilities
$|c_n|^2$), would require this fork of causality to  be replaced by a
fork of indeterminism (leading to different
{\em potential} states). The formal `plus' characterizing the
superposition would have to be replaced with an `or'. This discrepancy
represents the {\em quantum measurement
problem}. The  subsystem density matrices resulting from these two types
of fork are identical, since there is no way of distinguishing these
different situations operationally by a local observation. However, as
emphasized above, this argument does not {\em explain} the fork of
indeterminism that is at the heart of the probability
interpretation. If the pointer states
$\Phi_n$ are also orthogonal (as will have to be assumed for a
measurement), both sides of (13) are Schmidt-canonical.

This measurement problem exists regardless of the complexity  of the
measurement device (which may give rise to thermodynamically
irreversible behavior), and of the presence of
fluctuations or perturbations caused by the environment, since the states
$\Phi$ may be defined to describe this complexity completely, and even
include the whole `rest of the world'. The popular
argument that the quantum mechanical indeterminism
might, in analogy to the classical situation, be due to  {\em
distortions} (such as uncontrollable `kicks') during a
measurement (cf. Peierls 1985, for example) is
incompatible with a universal unitary dynamics. It
would require the existence of an initial  ensemble of microscopic
states that in principle had to determine the outcome. However, the
entropy of the rhs of (13) does not characterize
an ensemble as it would be required for this purpose.

If both systems in (13) are microscopic, the dynamics representing
the fork of causality can even be {\em practically} reversed  in order
to demonstrate that all arising components still exist. They
may then contribute to individual observable consequences that depend
on {\em all} of them, and on their relative phases. This excludes the
interpretation of (13) as representing a fork of indeterminism (a fork
between mere possibilities), even though von Neumann's fork of
causality is defined in a {\em classical} configuration space (in
terms of branching wave packets). If the transition from
quantum to classical were completely understood, it would have to explain
why the arena for the wave function {\em appears} as a space of `classical'
configurations in many situations.

   It
should be kept in mind, however, that the local  concepts of
relevance, such as
$\hat P_{\rm sep} , \hat P_{\rm local}$ and
$\hat P_{\rm classical} $, appear `natural' only to our classical
prejudice. In the unusual situation of EPR/Bell type experiments, quantum
correlations become relevant
even for  local observers. The locality of the {\em
dynamics},  in field theory described by means of point interactions, for
example,  merely warrants the dynamical consistency of this concept of
relevance, such as the approximate validity of autonomous	master
equations for
$\hat P_{\rm local}\rho $.

\section{Decoherence: Examples}

Using the terminology of Zwanzig projections introduced above,
decoherence may be defined as the dynamical justification of a {\em specific}
$\hat P_{\rm semidiag}$ for a given
system by presuming the relevance of the corresponding `local
perspective' that is formally represented by
$\hat P_{\rm sub}$.  If
this
$\hat P_{\rm semidiag}$ is valid under all normal circumstances, its
eigenspaces are labeled by classical properties
of the local (bounded but open) {\em systems}. In this way,
`quasi-classical' concepts (not just quasi-classical dynamics for
pressumed classical objects such as particles) {\em emerge} through
unavoidable interaction with the environment.

Equation (13) formulates the interaction of a microscopic system
$\phi$ with appropriate controllable `pointer states' $\Phi$  of a
measurement device. Its asymmetry in time  represents a
fact-like arrow (leading from factorizing to entangled
states). It can possibly be reversed (with sufficient effort) if both
subsystems are microscopic. For genuine measurements, the states
$\Phi_n$ must be quasi-classical. Assume now that the pointer
positions
$\Phi_n$ in (13) are replaced with uncontrollable (such as thermal)
states of the unavoidable environment of the system that is described by
the states
$\phi$, while this `system' may even represent the macroscopic pointer
states (being `measured' themselves by their
environment). Then this interaction cannot practically be reversed, and
measurements by means of macroscopic instruments can thus not be undone.
This (inter)action is  clearly analogous to Boltzmann's {\em Sto\ss
zahlansatz} in creating correlations which propagate away, while it
describes the specific quantum aspect of delocalizing phase relations
(`decoherence'). Its time arrow may therefore be referred to as
{\it quantum causality}. The resulting local (reduced) density
matrix $\rho_\phi$ in the sense of (5) describes an {\em apparent}
ensemble of quasiclassical states.

This interaction with the
environment, which leads to {\em ever-increasing entanglement}, is
practically unavoidable for most systems
in all realistic situations (Zeh 1970, 1971, 1973, Leggett
1980, Zurek 1981, 1982, Joos and Zeh 1985). It is this quantitative
aspect that seems to have been greatly overlooked when classical
systems were unsuccessfully described by a Schr\"odinger equation.
Decoherence is efficient, since it does not require the
environment to {\em act} on the system (as it would have to
do in order to induce `distortions' or Brownian motion). It does
neither depend on the concepts of momentum and energy,  nor include any
transfer of heat or the presence of an environmental heat bath, since
it may occur very far from equilibrium.

Decoherence is effective on a much
shorter time scale than thermal relaxation or
dissipation (see Joos and Zeh 1985, Zurek 1986). Its most
important message is that {\em there are no approximately
closed macroscopic systems} (save the whole universe). On the other
hand, systematic decoherence requires a `normal' environment that
monitors {\em certain} properties (represented by subspaces). The latter
may then appear as `classical facts', which exist regardless of
their observation, while their superpositions never occur
(locally). While this situation is the basis of Zurek's
(operationally understood) {\em existential interpretation} (Zurek
1998), it is here evidently derived from the assumption of an
`absolutely existing' universal wave function (of which it forms a
dynamically autonomous branch --- or a `consistent history' in terms
of wave functions). It appears indeed strange that most physicists
prefer to accept certain conceptual inconsistencies (often referred to
as `complementarity') which allow them to believe in the existence only
of what they can `see', even though their successful {\em and
consistent} theory tells them that there must be a lot more (unless
this theory is deliberately changed where it cannot be confirmed).

Some important
applications of decoherence will now be discussed.

\subsection{Trajectories}
In an imagined two-slit interference experiment with
`bullets' (or slightly more realistically with small dust particles or
large mole\-cules), not only their passage through the slits, but their
whole path would unavoidably be measured by scattered air molecules
or photons under {\em all} realistic circumstances. For macroscopic
objects we could simply confirm this fact by opening our eyes.
Therefore, no interference fringes could ever be observed --- even if
the resolution of our reading and registration devices were fine enough.
Macroscopic objects resemble alpha particles in a cloud chamber
(Mott 1929), since they can never be regarded as being isolated in a
vacuum (as it can be arranged for microscopic objects). Their unavoidable
entanglement with their environment leads to a reduced density matrix
that is equivalent to an
ever-increasing ensemble of narrow wave packets following slightly
stochastic trajectories.

For a continuous variable, such as the center of mass position of a
macroscopic particle, decoherence competes with the dispersion of  the
wave packet that is reversibly described by the Schr\"odinger
equation. Even the apparently small scattering rate of photons or
atoms off small dust particles in intergalactic  space
would suffice  to suppress all coherent spreading of the wave packets
(Joos and Zeh 1985). An otherwise free particle, for example, is then
dynamically described by the master equation
\be
i{{\partial \rho(x,x',t)}\over{\partial t}} = {1 \over{2m}} \left(
{{\partial^2}\over{\partial {x'}^2}}  -  {{\partial^2}\over{\partial
x^2}}
\right)
\rho - i\lambda (x-x')^2 \rho \quad ,
\ee
which can be derived from a
universal Schr\"odinger equation by assuming the future irrelevance  of
all initial correlations with the environment (cf. Joos' Chap.\ts3
and App.\ts1 of Giulini et al. 1996). The coefficient
$\lambda$ is here determined by the scattering rate and
its efficiency in orthogonalizing states of the environment. One does
not  have to {\em postulate} a fundamental semigroup in order to
describe  open quantum systems (Sect.\ts3). If the environment
represents a heat bath, (14) corresponds to the infinite-mass limit
of {\em quantum Brownian motion}  (cf.
Caldeira and Leggett 1983, Zurek 1991, Hu, Paz and Zhang 1992, Omn\`es
1997). This demonstrates that even for entirely negligible recoil (which
is responsible for noise and friction) there remains an important effect
that is based on quantum nonlocality. Although Brownian
fluctuations are required in a {\em thermal} environment, they describe
much smaller effects on the density matrix of macroscopic degrees of
freedom than decoherence.

Classical properties (e.\ts g. shape and position of a
droplet) thus {\em emerge}  from
the wave function
(and are maintained) in an irreversible manner.  Particle aspects
(such as tracks in a bubble chamber) are described by the reduced
density matrix because of unavoidable interactions with the
environment according to a universal Schr\"odinger equation. The
disappearance of interference between {\em partial waves} during a
{\em welcher Weg} experiment (Scully, Englert and Walther 1991,
D\"urr, Nonn and Rempe 1998) does not require a (fundamental)
wave-particle `complementarity'. Similarly, there is no superluminal
tunnelling (see Chiao 1998) in a consistent quantum description, since
{\em all parts} of a wave packet propagate (sub-) luminally, while its
group velocity does not represent propagation of a physical object in
the absence of a fundamental particle concept.

Master equations for open systems, such as (14),
can also be derived  by means of Feynman path integrals (Feynman and
Vernon 1963, Mensky 1979), that is, by using a {\em decoherence
functional}.  The path integral is here
used for calculating the propagation of wave functions of systems
together with their environments, while the latter are then dynamically
traced out. However, only because of the decoherence contained in this
procedure may  superpositions of different paths with their physically
meaningful phase relations (that is,  propagating {\em wave functions})
{\em appear} to a local observer as representing ensembles of
trajectories.

All
classical phenomena, even those representing `reversible' classical
mechanics, are  based on this (for all practical purposes)
irreversible decoherence, with a permanent production
of physical entropy that may be macroscopically
negligible, although it is large in terms of bits. These
`measurements' by the environment according to (13) must be
irreversible in this sense in order to avoid the possibility of
`quantum erasing' the
information from the environment, and thus to restore coherence (just
as for measurements proper).  {\em Recoherence} would mean that every
scattered particle were completely and coherently recovered in order
to  relocalize the initial phase relations. The terminology of quantum
erasing is therefore misleading: conventional erasing is understood as
the destruction (that is, dissipation) of information ---
corresponding to an increase of entropy ---, while the dissipation of
phase relations would just warrant perfect decoherence (rather than
undoing it). Experimental realizations of quantum erasers in certain
microscopic systems  (cf. Kwiat, Steinberg and Chiao 1992) do not
always {\em recover} the whole initial superposition --- they may
partly {\em rebuild} it.

\subsection{Molecular Configurations as Robust States}
Chiral molecules, such as right- or left-handed sugar,
represent another simple property controlled by decoherence. A
chiral state is described by a wave function, but in contrast to the
otherwise analogous spin-3 state of an ammonia molecule, say,
not by an energy eigenstate (see Zeh 1970, Primas 1983, Woolley
1986). The reason is that it is chirality (but not parity) that is
continuously `measured', for example by scattered air molecules (for sugar
under normal conditions on a decoherence time scale of the order
$10^{-9} sec$ --- see Joos and Zeh  1985). Measurements of the
parity of sugar molecules, or their preparation in
energy eigenstates, are therefore practically excluded, since this  would
require an even stronger coupling to the corresponding
device.

As a dynamical consequence, each individual molecule in a
bag of sugar must then retain its chirality, while a parity  state ---
if it had come into existence in a mysterious or expensive way ---
would almost immediately `collapse' into an apparent
mixture of both chiralities with equal
probabilities. Parity is thus not conserved for sugar molecules, while
chirality could  be confirmed `without demolition' when measured again.
(A mixture of chiralities would be {\em formally identical} to a mixture
of parities only in the pathological case of {\em exactly} equal
probabilities.)

This dynamical {\em robustness} of certain
properties under the influence of the environment seems to characterize
what we usually regard as `real classical facts' (in the  operational
sense), such as spots on the photographic plate, or any other `pointer
states' of a measurement device. The concept of robustness is
compatible with a (regular) time dependence, as exemplified in the
previous section for the center of mass motion of macroscopic objects.
 Since
entropy production by interaction with the environment is least for a
density matrix already diagonal in terms of robust states,
this property has been called a `predictability sieve', and
proposed as a {\em definition} of classical states (Zurek, Habib and Paz
1993).

Robustness also gives rise to
quasi-classical `consistent
histories' in terms of wave packets, and it is required for the
physical concept of memory, as in DNA, brains or computers --- with the
exception of quantum computers, which are extremely vulnerable to
decoherence (Haroche and Raimond 1996, Ekert and Jozsa 1996  --- see
also Zurek 1998). In contrast
to robust properties, which can be assumed to exist  as `facts'
regardless of their measurement, {\em potentially} measurable
quantities are  called `counterfactual' if they may occur in
superpositions. They must then not be assumed to possess definite
though unknown values. However, {\em all} situations can be described
and distinguished by means of decoherence in terms of a (f)actual
universal {\em wave function}.

Chemists know that atomic nuclei in large
mole\-cules  have to be described classically (for example by rigid
configurations, which may vibrate or rotate in a time-dependent
manner), while the electrons have to be represented by stationary or
adiabatically comoving wave functions. This asymmetry
is then often attributed to a Born-Oppenheimer approximation in terms of
the mass ratio. However, this argument is insufficient, since the
same approximation can be applied to {\em small} molecules for
calculating the stationary energy eigenstates with their discrete
energy bands. This insufficient argument is now also found in {\em
quantum gravity}, where it is claimed to explain classical spacetime
by merely employing a Born-Oppenheimer approximation with respect to
the inverse Planck mass. The argument cannot be improved by means a WKB
approximation for the massive variables, since this would still not
exclude broad wave functions (instead of narrow wave packets following
quasi-trajectories). The WKB approximation may only explain the
quasi-classical {\em propagation} of wave fronts
according to  geometric optics, and therefore the stability of wave
packets {\em once they have formed}.

The formation of quasi-classical wave packets for the
atomic nuclei in large molecules or for the gravitational field
can instead be explained by decoherence (Joos and Zeh
1985, Unruh and Zurek 1989, Kiefer 1992 --- see also Sect.\ts2.4). For
example, the positions of atomic nuclei in large molecules are
permanently monitored by scattering of (other) molecules. But why only
the nuclei (or ions), and why not even they in very small molecules? The
answer can only be quantitative, and it is based on a delicate balance
between internal dynamics and interaction with the environment, whereby
the density of states plays a crucial role (Joos 1984). Depending on the
specific situation, one will either obtain an approximately  unitary
evolution, a master equation (with time
asymmetry arising from quantum causality), or
complete freezing of the motion (quantum Zeno
effect).  The situation becomes  simple only for a `free'
massive particle, which is described by (14).

\subsection{Charge Superselection}
Gau\ss' law, $q = {1 \over 4 \pi}
\smallint {{\vec E} \cdot d{\vec S} }
$, tells us that every local charge is correlated with its Coulomb
field on a sphere at any distance. A
superposition of different charges,
\begin{eqnarray}
\sum_q{c_q \psi_q^{total} } = \sum_q {c_q \chi_q \Psi_q^{field} }
&=&  \sum_q {c_q \chi_q  \Psi_q^{near} \Psi_q^{far} } \nonumber \\
&=&:  \sum_q {c_q \chi_q^{dressed} \Psi_q^{far} }\quad , 
\end{eqnarray}
would therefore represent an entangled state of the charge and its
field. Here,
$\chi_q$  describes the bare charge, while
$\Psi_q^{field} \allowbreak = \Psi_q^{near} \Psi_q^{far}
$ is the wave functional of its complete field, symbolically written
as a tensor product of a near-field and a far-field.
The dressed (physical) state of a charge would then be described by a
density matrix of the form
\be
\rho_{local} = \sum_q { \vert \chi_q^{dressed} \rangle  \vert c_q\vert
^2 \langle
\chi_q^{dressed}
\vert }
\ee
if the states of the far field are mutually
orthogonal (uniquely distinguishable) for different charge $q$. The
charge is  decohered by its own Coulomb field, and no charge
superselection rule has to be {\em postulated} in a fundamental way (see
Giulini, Kiefer and Zeh 1995). The dressing of a charged particle by
its near-field (including reversible polarization of surrounding
matter) would {\em formally} decohere the bare charge, but remain
observationally irrelevant, since only the dressed particle can be
used in experiments.

While this result is satisfactory from a theoretical point  of view, a
more practical question is, at what distance and on what time scale a
point charge in a superposition of two different locations (such as an
electron during an interference experiment) would be decohered by the
corresponding dipole field. Classically, the retarded
Coulomb field on the forward light cone
would contain complete information about the path of the charge.
However, since interference between different electron paths has  been
observed over distances of the order of millimeters (Nicklaus and
Hasselbach 1993; see also Hasselbach's contribution to this
conference), one has to conclude that in QED Coulomb fields
contribute to decoherence by their monopole component
only.\footnote{This conclusion emerged from discussions with  E. Joos.}

This consequence, which appears surprising from a classical
point of view, may be readily understood in terms of quantized fields,
since photons with infinite wave length (representing static Coulomb
fields) cannot `see' position at all (even though their number may
diverge). Static dipole (or higher) moments do not possess any
far-fields, which are defined to decrease with ${1/ r^2}$ only.
Therefore, only the `topological' {Gau\ss}  constraint
$\partial_\mu F^{\mu 0} = 4\pi j^0$ remains
of the Coulomb field in QED. This requires that the
{\em observed} (retarded) Coulomb field is
completely described by transversal photons, corresponding to
the vector potential ${\vec A}$, with ${\rm div} {\vec A} = 0$ in the
Coulomb gauge, and in states obeying the {Gau\ss} constraint.
According to this picture, only the `positions' of electric field lines
--- not their total number or flux ---  represent dynamical variables
that have to be quantized. In this sense, charge decoherence has been
regarded as {\em
kinematical}, although it may as well be interpreted as being
dynamically {\em caused} in the usual way by the retarded Coulomb
field of the (conserved) charge in its past. However, the
absence of a dynamical Coulomb field, which may also eliminate the
need for renormalizing the mass of a charged particle by its Coulomb
field, is incompatible with the concept of a Hilbert space spanned by
direct products of {\em local} states.

Dipole moments (defining position {\em differences} of a point charge),
can thus be  `measured' either by the `real' emission of transversal
photons, or by the irreversible polarization of nearby matter
(K\"ubler and Zeh 1973, Zurek 1982a, Anglin and Zurek 1996). Emission
of photons requires {\em acceleration}. For  example, a {\em transient
dipole} of charge
$e$ and maximum distance $d$, existing for a time interval $t$,
involves accelerations
$a$ of at least the order $d/t^2$. According to Larmor's formula, the
intensity of radiation is at least ${2\over 3} e^2 a^2$. In order to
resolve the dipole, the radiation has to consist of photons with energy
greater than
$\hbar c/d$ (that is, wave lengths smaller than $d$).  The probability
that information about the  dipole is radiated away (by one photon, at
least) is then very small (of the order
$\alpha Z^2(d/ct)^3$, where $\alpha$ is the fine structure constant and
Z the charge number). In more realistic cases, such as interference
experiments with electrons, stronger accelerations may be involved,
although they would still cause negligible decoherence in most
situations.

The gravitational field of a mass point is similar to the  Coulomb
field of a charge. Superpositions of different mass should therefore
be decohered by the quantum state of spatial curvature, and give rise
to a mass superselection rule. However,
superpositions of slightly different mass evidently exist. They
form the time-dependent states of local systems, which would otherwise be
excluded. The quantitative aspects of this situation do not yet appear
to be sufficiently understood.

There would be no Coulomb field at all if the total
charge of the quantum universe vanished (cf. Giulini, Kiefer and Zeh
1995). The gravitational counterpart of this complete disappearance of a
classical quantity is the absence of time from a closed
universe in quantized general relativity (see Sect. 6.2 of Zeh 1999).


\subsection{Fields and Gravity}
Not only are quantum states of charged
particles decohered by their fields, quantum
field states may also be decohered in turn by the
sources on which they (re)act (see Kiefer's Chap.\ts4 of  Giulini et
al. 1996, and his contribution to these proceedings). In this case,
`coherent states', that is, Schr\"odinger's time-depen\-dent but
dispersion-free Gau\ss ian wave packets for the amplitudes of
classical wave modes (eigenmodes  of coupled oscillators), have been
shown to be robust for similar reasons as chiral molecules or the wave
packets describing the center of mass motion of quasi-classical
particles (K\"ubler and Zeh 1973, Kiefer 1992, Zurek, Habib and Paz
1993,  Habib et al. 1996). This explains why macroscopic states of
neutral boson fields usually appear as {\em classical fields}, and why
superpositions of macroscopically  different `mean fields' or
different vacua (see Sect.\ts6.1) have never been observed. In
particular, quantum (field) theory must not and need not be reduced to
a mere description of scattering processes with their unproblematic
probability interpretation in terms of  asymptotically isolated
fragments.

 These coherent (minimum uncertainty) harmonic oscillator states are
defined for each mode $k$ as the (overcomplete) eigenstates of the
nonhermitian photon annihilation operators $a_k$ with complex
eigenvalues
$\alpha_k$ (that is,
$a_k|\alpha_k \rangle = \alpha_k |\alpha_k\rangle$).  They represent
Gau\ss ian  wave packets centered at
a time-dependent mean field $\alpha_k(t)= \alpha_k^0 e^{i\omega t}$,
where
$Re(\alpha_k)$ and $Im(\alpha_k)$ are analogous to the mean position
and  momentum of a mechanical oscillator wave packet. Since the
Hamiltonian that describes the interaction with charged sources is usually
linear in the field operators $a_k$ or $a^\dagger_k$, these coherent
states form a robust `pointer basis' under normal conditions: they
cause negligible entanglement with their environment (here their
`sources').

In contrast to these superpositions of many different
photon numbers (that is, oscillator quantum numbers),
one-photon states resulting from the
decay of {\em different} individual atoms (or even the n-photon states
resulting from the decay of a different {\em number} of atoms) are unable
to interfere with one another, since they are correlated with
mutually orthogonal final states (different atoms
being in their ground state).  Two incoherent components of a
one-photon state may then appear (using Dirac's language) as `different'
photons, although photons are not even conceptually distinguishable
from another.  A coherent macroscopic (`collective') excited state {\em of
the source} would instead react negligibly (judged by means
of the inner Hilbert space product) when as a whole emitting a photon. It
would thus be able to produce the coherent (`classical') field states
discussed above (see Kiefer 1998).  In his textbook,  Dirac (1947)
discussed also states of {\em two} (or more) photons which are
entangled with one source state containing two or more decayed atoms
(and which may be described by a symmetrized product of one-photon
wave functions). Although these n-photon states form {\em coherent
components} of QED, their (two or more) {\em one-photon probabilities}
add again without interference (except for the exchange terms, which
give rise to the Hanbury Brown-Twiss effect).

Superpositions of {\em different} quasi-classical fields,
$c_1|\alpha_1\rangle+c_2|\alpha_2\rangle$, have  recently been produced
and maintained for a short time as one-mode laser fields in a cavity
(Monroe et al. 1996). Their smooth decoherence has also been
observed (Brun et al. 1996), as reported at this conference by
Haroche.

Similar arguments as used above for electromagnetic
fields apply to spacetime curvature in {\em quantum gravity} (Joos 1986,
Demers and Kiefer  1996 --- for applications to quantum cosmology see
Chap.\ts6 of Zeh 1999). One does not have to know its precise form (that
may be part of an elusive unified quantum field theory) in order to
conclude that the quantum states of matter and geometry (as far as this
distinction remains valid) must be entangled and give rise to mutual
decoherence. The classical appearance of spacetime geometry with its
fixed  light cone structure (that is presumed in
conventional quantum field theory) is thus no reason {\em not} to
quantize gravity. The beauty of  Einstein's theory can hardly be ranked
so much higher than that of Maxwell's to justify its exemption from
established theory. An exactly classical gravitational field interacting
with a quantum particle would be incompatible with the uncertainty
relations --- as is known from the early Bohr-Einstein debate. The
resulting density matrix (functional) for the gravitational tensor field
as a `quantum system' must therefore be expected to represent an apparent
mixture of different quasi-classical  curvature
states (to which even the observer is correlated).

Moreover, the
entropy and thermal radiation characterizing  a black hole or an
accelerated Unruh detector are
consequences of the entanglement between relativistic
vacua on two half-spaces separated by an event horizon  (see Gerlach
1988, and Sect.\ts5.2 of Zeh 1999). This entropy measures the same type
of `apparent' ensembles as the entropy produced according to the master
equation (14) for a macroscopic mass point. An event horizon need not
be different from any other quasi-classical property. Nonetheless, the
disappearance of coherence behind (even a virtual) horizon has been
regarded as a {\em fundamental} irreversible process of decoherence.
This does in no way appear justified.

\subsection{Quantum Jumps}
Quantum particles are often observed as flashes on a scintillation
screen, or heard as `clicks' of a Geiger counter. These macroscopic
phenomena are then interpreted as being caused by point-like objects
passing through the observing instrument during a short time interval,
while this is in turn understood as evidence for the discontinuous
decay of an excited state (such as an atomic nucleus). Using a {\em rate}
for stochastic `decay events' is equivalent to a master
equation. A {\em constant} rate would describe an
exponential decay law. Discrete quantum jumps
between two energy eigenstates have also been monitored for
single atoms in a cavity when strongly coupled to an observing device
(Nagourney, Sandberg and Dehmelt 1986, Sauter et al. 1986).

The Schr\"odinger equation, on the other hand, describes
a wave function (usually with angular distribution according to a
spherical harmonic) smoothly leaking out of a decaying
system (such as a `particle' in a potential well). This contrast
between observation and the Schr\"odinger dynamics
is clearly the empirical root of the probability
interpretation {\em in terms of discrete classical
concepts}, such as particles and events. Since its norm is conserved,  a
wave function can disappear exponentially only from a bounded open region
(usually an expanding sphere of size determined by the history of the
decaying object and the speed of the decay fragments). This wave function
represents a {\em superposition} (rather than an ensemble) of  different
decay times. Their interference and the dispersion of the outgoing wave
lead to further deviations from an exponential law.
Although they are too small to be observed in free decay,
they have been confirmed as
`coherent state vector revival' for photons emitted  into reflecting
cavities (Rempe, Walther and Klein 1987).

The appearance of `particles' as local
objects following tracks in a cloud chamber has been described in
Sect.\ts2.1 in terms of an apparent ensemble of
narrow wave packets. If droplets forming a condensate in the cloud
chamber, or similar phenomena, such as spots on a photographic plate and
clicks of a counter, appear at certain times, this is interpreted
as indicating `quantum events'. However, the same decoherence
which describes localization in space may, in the same sense, also
explain localization in time. Neither particles nor quantum jumps are
required as fundamental concepts (Zeh 1993). Whenever decay fragments (or
the decaying systems) interact strongly with their environment,
any interference between `decayed' and `not yet decayed' disappears on a
very short (though finite) decoherence time scale, similar to
Schr\"odinger's cat superpositions described in the previous section.
This time scale is in general far shorter than the time resolution in
genuine measurements. If decoherence is
even faster than relaxation into exponential behavior, decay may be
strongly suppressed (`quantum Zeno effect' ---
see Joos 1984 for a non-phenomenological discussion of its dynamics).

Decoherence thus leads to an
{\em apparent ensemble} of decay histories
consisting of a succession of events. The environment `monitors' the
decay status (in general uncontrollably) at all times with a resolution
defined by the decoherence time scale. A decaying system is then more
appropriately described by a decay rate than by a
Schr\"odinger equation. Its time dependence  would be exactly
exponential, while this master equation represents
only an approximate {\em local} consequence of the {\em global}
Schr\"odinger equation. Similarly, distinct decay energies forming an
initial superposition would usually be absorbed into mutually
orthogonal final states of the environment. Microscopic systems (with
their distinct energy levels) must therefore decohere into the
eigenstates of their Hamiltonian. This consequence of robust {\em
numbers} of emitted photons (Sect.\ts2.4) explains why stationary
states characterize the atomic world, and von Neumann spoke of an {\em
Eingriff} (intervention) required for their change.

It seems that this situation of continuously monitored decay has led
to the myth of quantum theory as a  stochastic
theory for fundamental {\em quantum events}
(cf. Jadczyk 1995). For example, Bohr (1928) remarked that ``the
essence" (of quantum theory) ``may be expressed in the so-called
quantum postulate, which attributes {\em to any atomic process} an
essential discontinuity, or rather individuality
\dots" (my italics). If this were true, there could be no lasers,
superconducters, or similar macroscopic superpositions.
Heisenberg and Pauli similarly emphasized that their preference for
matrix mechanics originated in its (as it now seems misleading)
superiority in describing discontinuities. However, according to the
Schr\"odinger equation and  recent experiments, the
underlying entanglement processes are smooth.
The short decoherence time scale mimics jumps between
energy eigenstates or, depending on the situation, into narrow wave
packets which in the Heisenberg-Bohr picture are interpreted as
{\em particles} (with classical properties restricted in validity by
the uncertainty relations in order to  comply
with the Fourier theorem).

While this new description may now appear as a consistent
picture in terms of wave functions, an  important  question remains open:
how do the probabilities which were
required to justify the concept of a density matrix in Sect.\ts1 have
to be understood if they do {\em not}  describe quantum jumps or the
spontaneous occurrence of classical properties through fundamental
`events'. These interpretational problems are discussed in Sect.\ts4.6
of Zeh (1999), but we have here to conclude that, from an external
point of view, the ensembles of wave
functions derived by decoherence are apparent ones.

\section{Quantum Dynamical Maps}
The phenomenological description of {\em open quantum systems} by
means of semigroups offers some novel possibilities which go beyond a
global Schr\"odinger equation. For example, quantum dynamical
maps have been used to formulate von Neumann's `first
intervention' (the reduction of the wave function) as part of the
dynamics (cf. Kraus 1971). This is possible, since semigroups cannot
only  describe the transition from pure states to
ensembles, but also the `selection' of
an {\em individual} member. Otherwise they are equivalent to an
entropy-enlarging Zwanzig-type master equation with respect to
$\hat P_{\rm sub}$  (or its
equivalent in terms of path  integrals --- Feynman and Vernon 1963).
Although the `irrelevant' correlations with the  environment, which
would arise according to the exact global formalism, represent quantum
entanglement, they are here usually not distinguished from
classical statistical correlations (defined for ensembles only) when it
comes to applications. This attitude is equivalent to a popular but
insufficient `naive' interpretation of decoherence, which  pretends
to derive {\em genuine} ensembles.

Quantum theory is sometimes even {\em defined} as
describing open systems by means of a dynamical semigroup, that  is,
as a time-asymmetric local statistical theory. (Hence the term
`statistical operator' for
the density operator.) However, this `minimal statistical interpretation'
is insufficient as a fundamental theory, as it neglects the essential
difference between genuine and apparent ensembles,
and thus all  consequences of entanglement
beyond the considered systems (quantum nonlocality). The superposition
principle has even been claimed to be
derivable (cf. Ludwig 1990), although it must then be
re-introduced in a different way (for example by changing the laws of
statistics --- in conflict with any ensemble interpretation).

Semigroups are certainly
mathematically elegant and powerful. Therefore, they would represent
candidates for {\em new} (fundamental) theories if conventional
(Hamiltonian) quantum theory should prove wrong as a universal theory.
The question is whether mathematical elegance here warrants physical
relevance or is merely
convenient within a certain approximation. To quote Lindblad (1976): ``It
is difficult, however, to give physically plausible conditions
\dots \ which rigorously imply a semigroup law of motion for the
subsystem.
\dots \ Applications \dots \ have led some authors to introduce  the
semigroup law as the fundamental dynamical postulate for open
(non-Hamiltonian) systems." Such a law would {\em fundamentally} introduce
an arrow of time, but it
would depend on the choice of systems (and in some cases
contradict experiments that have already been performed).

The simplest quantum systems (such as spinors) are described by a
two-dimensional Hilbert space. Their density matrix may be
written by means of the Pauli matrices $\sigma_i$ ($i = 1,2,3$) in
the form
\be
\rho = {1 \over 2}(1 + {\vec \sigma} \cdot {\vec \pi}) \quad ,
\ee
where the (mathematically) real {\em polarization vector} ${\vec \pi} =
{\rm Trace}\{{\vec \sigma}\rho \}$ --- that is, the expectation value of
{\em all} spin
components --- completely defines $\rho$ as a general hermitian
$2 \times 2$ matrix of trace 1. The latter is in turn equivalent to a
(genuine or apparent) {\em ensemble} of orthogonal states (a spinor
basis). The length of
${\vec \pi}$ is  a measure of purity, since
${\rm Trace} \{ \rho^2 \} = (1 + {\vec \pi}^2)/2$, with ${\vec \pi}^2 \le 1$.
A pure state corresponds to a unit polarization vector, while an
arbitrary density matrix (a general `state' in the language of
mathematical physics) is characterized by the mean value
$\vec \pi =
\sum_\alpha p_\alpha {\vec \pi}_\alpha$ of all unit vectors ${\vec
\pi}_\alpha$ in an  ensemble of spinors that may represent this density
matrix.

A general trace-preserving linear operator
$\hat P$ on $\rho$ must be defined on 1 and ${\vec \sigma}$ in order
to be completely defined:
\be
\hat P 1 := 1 + {\vec \pi}_0 \cdot {\vec \sigma} \quad \quad \quad
\hat P {\vec \sigma} := {\vec A} \cdot {\vec \sigma} \quad ,
\ee
with a real vector ${\vec \pi}_0$ and a linear vector transformation
${\vec A}$.
$\hat P$ is idempotent (a Zwanzig `projector')  if  ${\vec A}^2 =
{\vec A}$ and
${\vec \pi}_0 \cdot {\vec A} = 0$ (${\vec A}
= 0$, for example). If ${\vec \pi}_0 \ne 0$, $\hat P$ creates
information --- even from  the unit matrix.

Dynamical combination of the projection $\hat P$ with a Hamiltonian
evolution  (rotation of
${\vec \pi}$) in the form of a master equation  leads to
the {\em Bloch  equation} for the vector
${\vec \pi}(t)$,
\be
{d {\vec \pi} \over dt}= {\vec \omega} \times  ({\vec \pi} -
{\vec \pi}_0 ) - \sum_i
\gamma_i (
\pi^i  -  \pi_0^i ) {\vec e}_i
\ee
in a certain vector basis $\{ {\vec e}_i \}$ (cf. Gorini, Kossakowski
and  Sudarshan 1976). Values of  $\gamma _i < 0$ or $|{\vec \pi} _0 | >
1$ would violate the positivity of the density
matrix\ts
at some
$t > 0$ (cf. Sect.\ts4.2), and have thus to be excluded.\footnote{As
mentioned before, {\em all} subsystem density matrices remain positive
under a global Hamiltonian dynamics, and even under a collapse of the
global state vector. This property of `complete positivity'
has to be separately {\em postulated} for phenomenological quantum
dynamical maps (cf. Kraus 1971), thus further illustrating that
these maps do not describe a {\em fundamental} quantum
concept.} The second term on the rhs describes anisotropic damping
towards
${\vec \pi}_0$. This formation of new information may describe very  different
situations --- for example equilibration with a stationary  external
heat bath of given temperature, or evolution towards a certain
measurement result. However, hermiticity of
$\hat P$ (corresponding to a genuine projection operator) would
require
${\vec \pi}_0 = 0$ and
${\vec A} = {\vec A}^{\dag} $, that is, a projection of
vectors ${\vec \pi}$ in space.

If the two-dimensional Hilbert space describes something else than
spin, such as isotopic spin or a $K, \bar K$ system, the polarization
vector lives in an {\em abstract} three-dimensional
space, with environmental conditions that cannot practically be
`rotated'. The abstract formalism can also be generalized to
$n$-dimensional Hilbert spaces. For this purpose the Pauli matrices
have to be replaced with the
$(n^2 - 1)$ hermitean generators of $SU(n)$, while the real `coherence
vectors' (the generalizations of the polarization
vector ${\vec \pi}$) now live in the vector space spanned by them. For
example, SU(3) gives rise to the `eight-fold way'. The most important
difference is that there are now more than one (in fact,
$n-1$) {\em commuting} hermitian generators. They may contain a
nontrivial  subset that is decohered under all realistic
environmental conditions, and thus may form the center of a
phenomenological set of observables (the set of `classical observables'
--- cf. Sect.\ts2).

In the infinite-dimensional Hilbert space of quantum
mechanics, the {\em Wigner function}
\begin{eqnarray}
&W&(p,q) := {1 \over \pi} \int e^{2ipx} \rho(q+x,q-x)\, dx  \nonumber
\\
 &\equiv&
{1 \over {2\pi}} \int \int \delta \left( q-{{z+z^\prime}\over
2}\right)  e^{ip(z-z^\prime)} \rho(z,z^\prime)
\, dz dz^\prime =: {\rm Trace} \{ \Sigma_{p,q} \rho \}
\end{eqnarray}
(written in analogy to ${\vec \pi} =
{\rm Trace}\{ {\vec \sigma}\rho\}$)
assumes the role of the Bloch vector. Evidently,
$\Sigma_{p,q}(z,z^\prime)  :=
{1 \over {2\pi}} e^{ip(z-\nobreak z^\prime)}\, \delta (q-\nobreak
{z+z^\prime\over 2})$ is the generalization of the Pauli matrices
(with `vector' index
$p,q$).\footnote{On a finite
interval of length $L$, $\Sigma_{p,q}$ would require an
additional term $ -{1
\over {2\pi L}} \exp [ip(z-z^\prime)]$ in order to
remain traceless. In (20), $W(p,q)$ is then accordingly replaced
with
$W(p,q) - {1 \over L} \int W(p,q)dq$ as the generalized Bloch vector.
(Note that in Zeh (1999) the factor
$\exp [ip(z-z^\prime)] $ of this additional $1 \over L$-term has
erroneously been replaced by the $\delta$-function from (20).)}
Therefore, the Wigner function is a continuous set of  expectation
values, which form the components (one for each point in phase space)
of a generalized coherence vector.  This `vector' of expectation values
characterizes the density matrix
$\rho$ again completely, and regardless of its interpretation
according to Sect.\ts1. It does neither represent a quantum state
nor a probability distribution on phase space (as is evident from its
possibly negative values), even though it allows one to calculate all
expectation values in the {\em form} of an ensemble mean,
$<F> = \smallint f(p,q)W(p,q)dpdq$.

Lindblad (1976) was able to generalize  the Bloch equation to
infinite-dimensional Hilbert spaces. He wrote it (in its form applicable
to the density matrix) as
\be
i{{\partial \rho} \over \partial t} = [H,\rho] - {i \over 2} \sum_k
\left(  L^{\dag}_k L_k \rho + \rho L^{\dag}_k L_k - 2L_k  \rho
L^{\dag}_k
\right) \quad ,
\ee
with arbitrary generators $L_k$ in Hilbert space. It represents
creation (localization) of information in the considered system, that is,
a decrease of the corresponding von Neumann entropy (such as
described by
${\vec \pi}_0$ in (18)), precisely if some generators do not  commute
with their hermitian conjugates
$L^{\dag}_k$. This can be demonstrated by applying the non-Hamiltonian
terms of (20) to the unit matrix $\rho = 1$. Otherwise it describes
information {\em loss} (a genuine Zwanzig projection). This can also be
seen from the general representation of a Zwanzig projector in quantum
mechanical Hilbert space,
$\hat P \rho = \sum_k V_k \rho V_k^{\dag}$, which is
analogous to the square root of a positive operator in its eigenbasis
for $V_k = V_k^{\dag}$. If
$L^{\dag}_k = L_k$, the Lindblad terms can be written in the form  of a
double commutator,
$L^2 \rho + \rho L^2 - 2 L
\rho L = [L,[L,\rho]]$. For $L = \sqrt{2\lambda}x$ one recovers
(14), that is, decoherence in the $x$-basis, as it could
be {\em derived} from unitary interaction with the environment (and shown
to be practically unavoidable for macroscopic variables).

One may similarly describe other `unread measurements' and their
corresponding loss of phase relations. However,
`damping' towards a {\em pure} state (a semigroup proper) according to
the second term of (17) with a unit vector ${\vec \pi}_0$ allows one even
to describe dynamically the  transition from the initial state vector
into a (freely chosen) definite measurement outcome (a
`collapse' --- in contrast to a local
{\em or global} Schr\"odinger dynamics). This can then readily be
combined with a stochastic formalism
representing an appropriate dice (or random number generator) that
selects  pure states according to the Born-von
Neumann probabilities (Bohm and Bub 1966, Pearle 1976, Gisin 1984,
Belavkin 1988, Di\'osi 1988). If applied continuously, such as by means
of the
It\^o process, this formalism
describes measurements
phenomenologically as a smooth indeterministic process (that does not
distinguish between ensembles and entanglement).

Many explicit models have
been proposed in the literature (see Stamatescu's Chap.\ts8 of Giulini
et al. 1996). Some of them merely replace the  apparent
ensemble arising through
decoherence for a bounded open system with a {\em genuine} one (Gisin and
Percival 1992). The system  is then assumed always to possess its own (yet
unknown) state
$\psi(t)$ that follows an indeterministic trajectory in its Hilbert
space according to a quantum Langevin
equation --- in conflict with the exact global dynamics that leads to
entanglement. Therefore, this  `quantum state diffusion model'  is
essentially equivalent to what I have called above the `naive
interpretation' of decoherence. It may  serve as an  intuitive picture
(or tool) for many practical purposes {\em if} (and insofar as) it
selects the dynamically  robust wave packets described in Sect.\ts2.2.
However, it would be severely misleading if this formalism (based on
the concept of a density matrix) gave rise to the impression of {\em
deriving} a real collapse by just taking into account the interaction
with the environment. If real physical states are described by wave
functions, there are only two possibilities: deviations from the
Schr\"odinger equation or the Everett interpretation.

Many contributions in the literature
remain ambiguous about their true intentions, or simply disregard the
difference between genuine and apparent ensembles
(proper and improper mixtures).  In particular,
the quantum state diffusion model is {\em not} appropriate to define a
fundamental dynamical process, since the resulting pure states
$\psi(t)$ of a system would in general not define states for any of
its subsystems (which could as well have been chosen as {\em the}
system, and thus have led to a different stochastic evolution). The
picture of a trajectory of states
$\psi(t)$ for a macroscopic system that is not the whole universe is
simply in drastic conflict  with quantum nonlocality.

Other models therefore attempt to reproduce
the observed statistical aspects of quantum theory (as they
occur in measurements, for example) by means of dynamical
laws which may be truly fundamental. Since they cannot remove all
entanglement, they cannot describe trajectories of wave functions
$\psi(t)$ for all `systems'. Measurements are special
applications of this general stochastic quantum
dynamics that describes an increase of ensemble
entropy. Explicit modifications of a {\em universal} Schr\"odinger equation
were originally suggested in the form of stochastic `hits', assumed to act
in addition to the unitary evolution in order to suppress coherence with
growing distance (Ghirardi, Rimini and Weber 1986). They were postulated to
occur rarely for individual particles, but sufficiently often for entangled
aggregates of many particles in order to describe
quasi-classical behavior. This proposal was later
formulated as a continuous process as indicated above (Pearle 1989,
Ghirardi, Pearle and Rimini 1990). Since these models lead to novel
predictions, they can be distinguished from a universally
valid Schr\"odinger equation. In their original form they would either be
completely camouflaged by environmental decoherence
(Joos 1986, Tegmark 1993), or are ruled out by existing experiments
(Pearle and Squires 1994). They cannot be excluded in general, however,
if one allows them to occur clearly {\em after} environmental
decoherence has occurred in the observational chain of
interactions. Their precise form would then be hard to guess in the
absence of any empirical hints.

Several authors have suggested to find the root of a fundamental
quantum indeterminism in gravity. Their main motivation is
the apparently classical nature of spacetime curvature. However, it has
been indicated in Sect.2.4 that spacetime need not be classical.
Collapse  models along these lines have
been proposed in a more or less explicit form (Penrose 1986,
K\'arolyh\'azy, Frenkel and Luk\'acz 1986). They regard the quantum state
of the gravitational field either as an environment to matter in a
specific quantum state diffusion model (as
criticized above --- see Di\'osi 1987), or they are using an arising
event horizon as a `natural' boundary to  cut off
entanglement (Hawking 1987, Ellis, Mohanty and
Nanopoulos 1989). Even though this boundary between `systems' may appear
natural, this procedure would still not define an objective fundamental
process (see also Myers 1997). In particular, the horizon depends on the
complete history of motion of the observer. Under no circumstances would
this proposal justify the replacement of apparent ensembles with genuine
ones, unless explicitly postulated so in an invariant form as a
{\em modification} of unitary quantum dynamics.

\end{document}